\documentclass[aps,twocolumn,pre,showpacs,superscriptaddress]{revtex4-1}
\usepackage{amsmath,amssymb}
\usepackage{graphics,graphicx}
\usepackage{dcolumn,bm}
\usepackage{psfrag}
\newcommand{\cu}
{\affiliation{Department of Physics, University of Calcutta,
92 Acharya Prafulla Chandra Road, Kolkata 700009, India.}}

\begin{document}

\title{Condensation Transition in a conserved  generalized interacting  Zero Range Process}

\author{Abdul Khaleque}%
\cu
\author{Parongama Sen}%
\cu

\begin{abstract}
A  conserved  generalized zero range process is considered 
in which
two sites  interact such that 
particles hop    
 from the more populated site to the other  with a probability $p$. 
The steady state particle distribution function $P(n)$ is obtained using both analytical and numerical methods. 
The system goes through several phases as $p$ is varied. 
In particular, a  condensate phase appears  for $p_l < p < p_c$, where the bounding values 
depend on the range of interaction, with  $p_c <  0.5$ in general.  
Analysis of $P(n)$ in the condensate phase  using a known scaling form  
shows there is universal behaviour in the short range process while the infinite range process displays non-universality. 
In the non-condensate phase above $p_c$, two distinct 
regions are identified:  $p_c < p \leq  0.5$ and $p>  0.5$; a scale emerges in the system in the latter and this feature is present for all ranges of interaction. 

\end{abstract}

\pacs{05.40.-a,05.70.Ln,05.70.Fh,02.50.-r}

\maketitle

\section{Introduction}

Condensation in various stochastic processes have attracted a lot of attention recently. Jamming in traffic flow~\cite{Loan,chowdhury,kaup},
 pathological phases in quantum gravity~\cite{Bialas}, condensation
of edges in complex networks~\cite{krapivsky,bian}, wealth condensation in microeconomics~\cite{bouch,budra} and clustering in shaken 
granular gases~\cite{van} are some well known examples.
 In the  condensate phase, a macroscopic fraction of particles  lies in a small region of the system.
A non-equilibrium system in which  condensation occurs in real space is the mass transport model where mass accumulates locally in the condensate phase~\cite{satya,zia,Waclaw}.   
The zero-range process (ZRP) which  was  introduced in~\cite{spitzer} as a mathematical model of  interacting particle system is  a special case of the mass transport model. 
It has been one of the  most studied models in the literature of non-equilibrium
phenomena  in recent years~\cite{kafri,evans1,evans2,gross,godr,jeon,gross,godr,ferrari,arm1,arm2}. 

In the  ZRP, particles hop from one site to another on a one dimensional chain of $L$ sites and the hop rate usually depends
only on the occupation number of the departure sites in the non-interacting case.
The particle number $N$ is  given by $\rho L$
where $\rho$ is the density and constant in a conserved system. 
A condensation can 
occur if  $\rho$  exceeds
a critical value $\rho_c$ \cite{evans2}, mostly studied in the asymmetric case. In the condensate phase, all excess particles accumulate on a single site.
This is exhibited in the form of a  bump occurring near  the tail of the particle distribution function. 
In the cases considered so far, the departure and destination sites are  fixed and are nearest neighbours apart from some exceptional cases and on graphs~\cite{Waclaw3}.

In this paper we have  considered  a modified  model with conservation (i.e. particle number is fixed),
 where a pair of sites  are chosen between which a particle transfer 
takes place. 
However, which one will be the departure/destination site depends on the 
occupation number of both sites. 
 Precisely, we consider that the hop will take place from the more populated site to the other with probability $p$  and in the reverse 
direction with probability $1-p$.
If the two interacting sites are equally populated, a hop takes place randomly with equal probability from either of them. 
Also, if a site is empty, there is a particle transfer from the occupied site with probability $p$ while there is no question of a reverse transfer.
Interactions have been incorporated 
in  some modified ZRP and   mass transport models 
\cite{evans2,satya97,coco,luck,Waclaw1}  
where    hopping probability   depends on the 
 occupation numbers of two or more  sites.
Our model is clearly different from such cases as well  as from the non-interacting models.
We call this a generalized interacting zero range process; the $p=1/2$ limit corresponds to  the conventional zero range process.

In section II, we describe the general features of the model and a mean-field  approach is discussed in section III. Results using numerical 
simulations are 
presented in section IV.   In the last section, a short summary and some discussions have been made.

\section{The model: Limits and general features} 
 \label{model}
\begin{figure}[!h]
\includegraphics[width=8cm,angle=0]{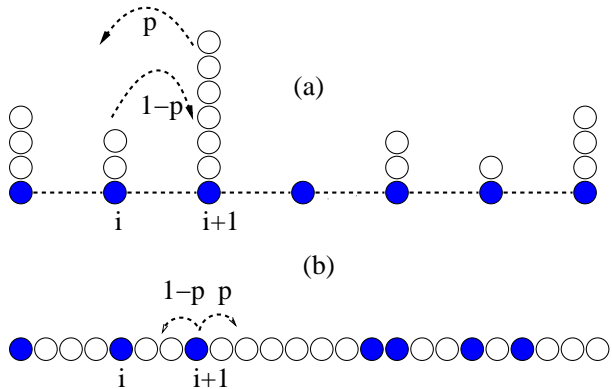}
\caption{Mapping between (a) the short range ZRP and (b) the traffic model. (a) Sites are shown by filled circles and particles by open circles.
Here a particle hops from  the $i+1$th site which is more populated 
 to the $i$th site with probability $p$  and  from the $i$th site to $i+1$th site with probability $1-p$. (b) The corresponding
  traffic problem in one dimension, where the occupation number
corresponds to the number of empty sites (open circles) to the right of the vehicles (filled circles) and the motion is towards the neighbour located further with probability $p$.
 \label{zrp}}
\end{figure}

The present model is inspired by its analogy to certain physical problems. In case the hops occur between nearest neighbours 
(short range (SR) case), it can be easily mapped to a traffic problem with hard core repulsion, where particles can move either way.   
The equivalent dynamical rule is movement towards the neighbour located further with probability $p$ (Fig. \ref{zrp}). 
Note that this is simply opposite to the dynamics considered in \cite{spp} where particles move towards their 
nearest neighbour (and annihilate when they collide). In the model considered in this paper,
for $p=1$,  the particle will always travel away from its nearer neighbour so as to avoid a collision.
The $p=1$ ($p=0)$ limit (in the SR model) can also be regarded as  a  system of identically (oppositely) charged particles. 
We considered also the long range  (LR) version of the model where the pair of interacting sites are chosen randomly.

We intend to obtain the particle distribution function $P(n)$ as time $t \to \infty$, i.e., the probability that there are 
$n$ particles on a site in the steady  state for different values of $p$ 
with   a fixed value of $\rho$.


For the  limiting case $p=1$, there will be hops to lesser populated sites always.  A steady state
will correspond to all sites having identical  number of particles equal to $\rho$ ($\rho \geq 1$) if 
there were no transfers between equally populated sites.
 However, since we allow transfers even in this case, 
 one may expect the probability $P(n) $ to be peaked at $n = \rho$ with a finite width.   
 On the other hand,  
for small $p$, there will be more transfers from lesser populated sites leading to a tendency of accumulation
of particles on a single  site which corresponds to a condensate phase. Hence 
one can expect a transition between the condensate and non condensate phase at a critical value of $p$. 
The condensate-non-condensate phase transition, if any, is qualitatively 
similar to the jamming transition in the traffic model; the condensate phase corresponding to the jammed phase.
Thus the non-condensate phase is also termed as the fluid phase. 

In  general, for   mass transport models   it was found that in the condensate state, 
  $P(n)$ can be analysed using the scaling form \cite{satya}:
\begin{equation}
 P(n)  \propto \frac{1}{L^{\gamma+1}}f[\frac{n-M}{L^\gamma}],
\label{scale}
\end{equation}
where $M$ is related to the excess mass. Our goal will be to find out whether a similar scaling form is valid for the present case and the 
value of the exponent $\gamma$ in the condensate phase.
Results are  essentially independent of the value of $\rho$ apart from trivial scaling factors
and presented  for $\rho = 5$ for all cases.   
\section{Mean-field approach}
 \label{mean}
The long range model can be regarded as a mean field model for which a master equation for $P(n)$ 
  can be written as
\begin{widetext}
\begin{eqnarray}
 \frac{dP(n,t)}{dt}&=& -p\sum_{m<n} P(n,t) P(m,t)-(1-p)\sum_{m>n} P(n,t) P(m,t) \nonumber \\
& & -p\sum_{m>n} P(n,t) P(m,t)-(1-p)\sum_{1\leq m<n} P(n,t) 
  P(m,t)+p\sum_{m<n+1} P(n+1,t) P(m,t)+ \nonumber \\
& & (1-p) \sum_{m>n+1} P(n+1,t) P(m,t)+p\sum_{m>n-1} P(n-1,t) 
 P(m,t)+(1-p)\sum_{1 \leq m<n-1} P(n-1,t) P(m,t) \nonumber \\
&& -2 [P(n,t)]^2+[P(n+1,t)]^2
+[P(n-1,t)]^2
\label{master1}
\end{eqnarray}
for $n\geq 1$ and 
\begin{eqnarray}
 \frac{dP(0,t)}{dt}&=& -p\sum_{m>0} P(0,t) P(m,t)+p P(1,t) P(0,t) +
(1-p)                
 \sum_{m>1}P(1,t) P(m,t)+[P(1,t)]^2
\label{master2}
\end{eqnarray}
 for $n=0$. 
\end{widetext}
Here the transition rates have been assumed to be unity and $t$ taken to be a dimensionless variable. 
The upper limit in the summations is  $\rho L$ unless otherwise restricted.
The conditions to be satisfied by $P(n,t)$ are $\sum_n P(n,t) = 1$ (normalisation) and $ \langle n \rangle = \sum_n n P(n,t) = \rho$ (conservation)
at all times.  We have used the notation $P(n, t \to \infty) = P(n)$.

\begin{figure}[!h]
\includegraphics[width=8cm,angle=0]{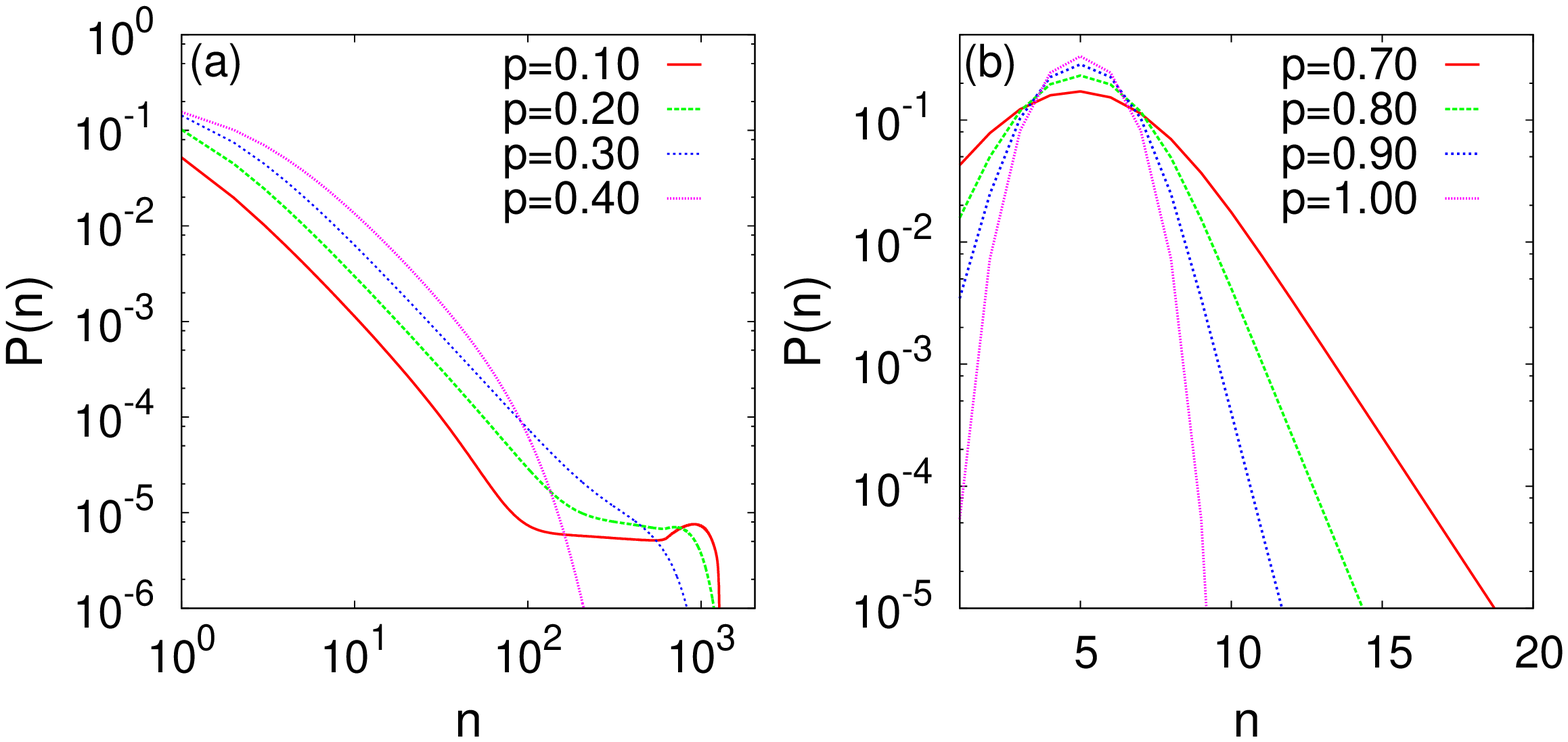}
\caption{ME: Occupation probability distribution $P(n)$  
for (a) lower values of  $p$ and (b) higher values of  $p$.  System size is $L=256$.
 \label{prob_eqn}
}
\end{figure}
Solving  these equations numerically, we calculated  $P(n)$,  the steady state value (see Fig.~{\ref{prob_eqn}).
The initial values  $P(n,t=0)$ are taken to be completely random.
For lower  values of $p$ a  bump  appears
at  $n={\mathcal {O}}(\rho L $) indicating the presence of a condensate. 
A systematic study of the distribution function for  small $p$,    
however, is difficult as the tail
of the distribution takes very long time  to equilibrate and this leads to   accumulation of  numerical errors
in solving the differential equations. 
At small $p$, the bump  changes shape as the observation time is increased while for small $n$, $P(n)$ remains more or less constant even at later times. 
Although the evidence of the transition  is present in the results as
the  bump disappears for larger values of $p$, the ME results are  less reliable for small $p$ and it 
is not possible to locate the  transition point very accurately.
Nevertheless, one can check whether eq. (\ref{scale}) is valid here by plotting the rescaled probabilities for
different values of $L$ and achieve a data collapse using trial values of $\gamma$. In Fig.~\ref{scale_eqn}, 
the data for $p = 0.05$ are shown where we see that with $\gamma = 0.99 \pm 0.005$,     
the data collapse is fairly appreciable.

\begin{figure}[!h]
\includegraphics[width=6cm,angle=0]{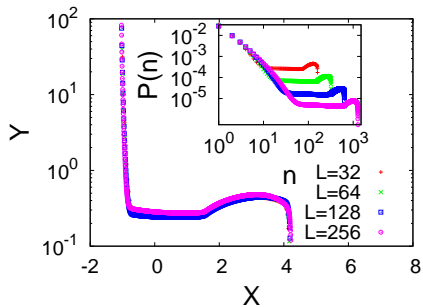}
\caption{ME: Scaling collapse of occupation probability distribution $P(n)$,
 where $Y=P(n) L^{\gamma+1}$ has been plotted against $X=\frac{(n-L)}{L^{\gamma}}$. 
 Inset is for the occupation probability distribution $P(n)$  for different  system sizes.  
Particle density is $\rho=5$ and $p=0.05$.
 \label{scale_eqn}
}
\end{figure}

\begin{figure}[!h]
\includegraphics[width=8cm,angle=0]{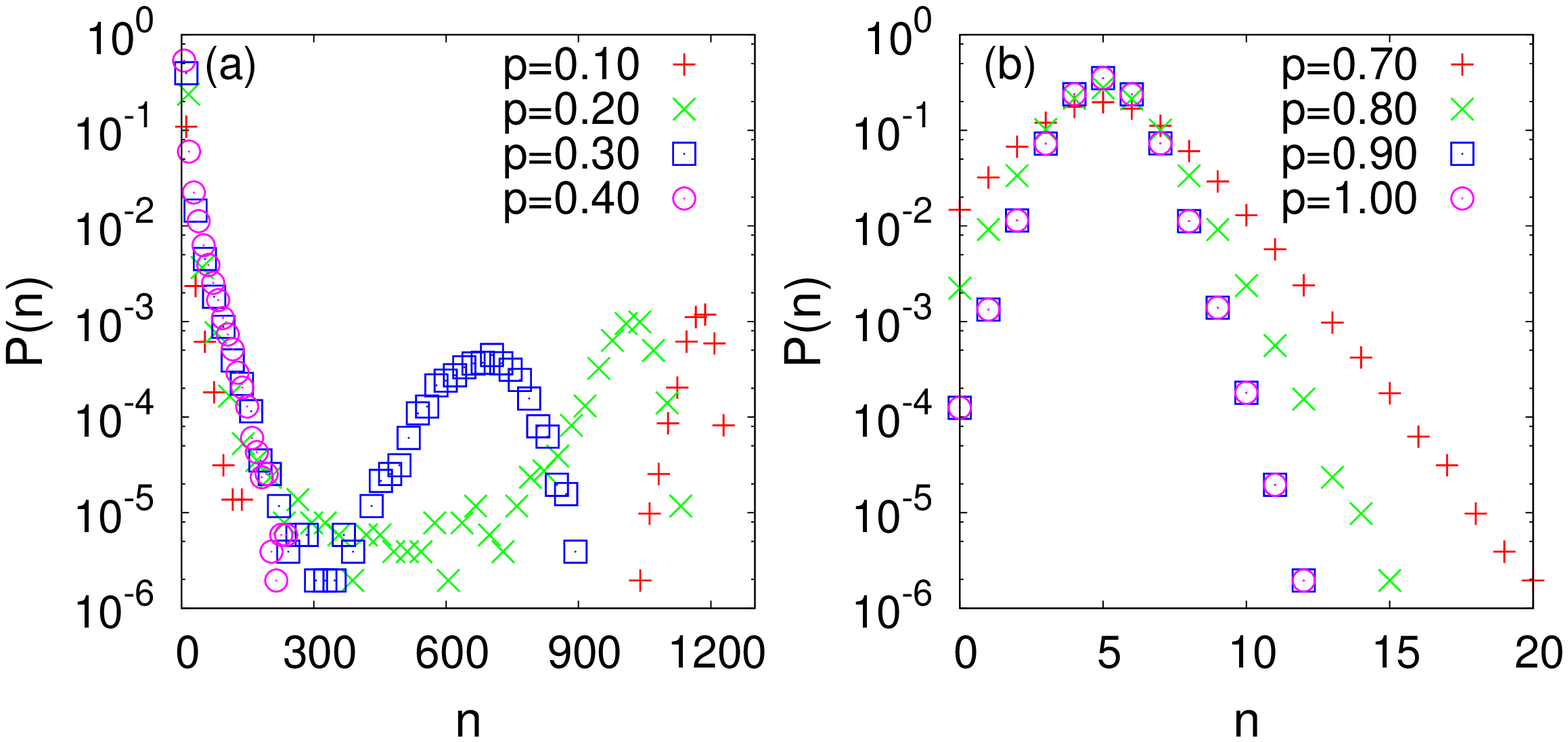}
\caption{LR:  Occupation probability distribution $P(n)$ is for (a) lower values of  $p$ and (b) higher values of  $p$.  System size is $L=256$. 
 \label{lr_prob_dist}
}
\end{figure}

\begin{figure}[!h]
\includegraphics[width=8cm,angle=0]{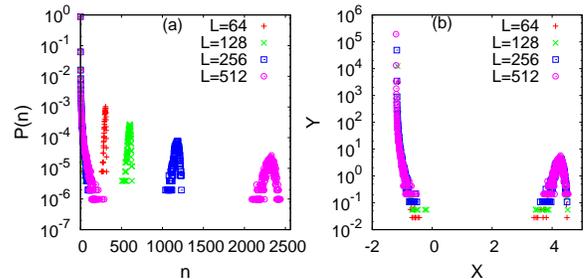}
\caption{LR: (a) Occupation probability distribution $P(n)$ for different  system sizes  for $p=0.10$ 
and (b) the scaling collapse of the same. Here $Y=P(n) L^{\gamma+1}$ has been plotted against $X=\frac{(n-L)}{L^{\gamma}}$. 
Particle density is $\rho=5$.
 \label{lr_prob_scale}
}
\end{figure}

\section{Results using numerical simulations}
 \label{result}
In this section we discuss results obtained using numerical simulation considering both the long range and short range models.
In general, we have taken systems with size $L\leq 1024$ and taken averages over  at least two thousand configurations. Initially, all the sites are randomly occupied.
Asynchronous dynamics have been used to update the state of the system.

\subsection{Condensate to fluid phase transition}
\subsubsection{Long Range (LR) case}
As the master equation (ME) approach poses difficulty to analyse the tail of the distribution for small $p$, 
we  used  Monte Carlo simulation to calculate $P(n)$. 
The steady state results for  LR case 
show clearly the presence of the  bump right from $p=0$ up to a value $p=p_c$ (see Fig. \ref{lr_prob_dist}(a)). 
Simple inspection shows that the transition point $p_c$ beyond which the bump vanishes 
is close to $p=0.33$. 
It is possible  to analyse $P(n)$ for $p < p_c$ and do a scaling analysis using data for 
different values of $L$. We find that  indeed the  scaling form given by (1) is valid with 
$M = cL$ as the data for different system sizes collapse to a single curve when an appropriate value of $\gamma $ is used (see Fig. \ref{lr_prob_scale}). 
The proportionality constant $c$ 
is in general 
equal to $1$ 
 to obtain  a good collapse.
However, there are some intricacies involved in this choice which is discussed later in section \ref{summary}. 

The value of $\gamma$  
 (Table  
\ref{tab:expo}) shows a slow decrease with $p$.  
This implies  the LR case has a non-universal behaviour. 
This non-universality is presumably related to the fact that the location
of the  secondary peak in $P(n)$ depends on $p$ as shown in  Fig. \ref{lr_prob_dist}(a).

The estimate of the transition point can be more accurately made from the calculation of fluctuations.
In the condensate phase, $P(n) $ has considerably large  values  only for $n$ close to zero  and $\rho L$.
As an approximation, one can take  $P(n)$ to be   a double delta function at these two values,  
such that 
\begin{equation}
P(n) = A\delta(n) + B\frac{1}{L^{\gamma+1}}\delta ( \frac{n-cL}{L^\gamma}). 
\end{equation}
The constants $A$ and $B$ can be found out from the  conditions   
$\int P(n)dn = 1$ (normalisation condition) and $\int n P(n) dn  
 = \rho$ (fixed average value). This gives
  
\[
A+ B/L= 1
\]
and 
\[
Bc = \rho.
\]
In the thermodynamic  limit  $B/L$ vanishes giving  $A = 1$. Using these values, 
the  fluctuation  $ \sqrt {\langle \Delta n^2}\rangle =\sqrt{\langle n^2\rangle- {\langle n \rangle }^2}$  equals $\sqrt {\rho c L - \rho^2} $. Hence to the leading order,  $ \sqrt {\langle \Delta n^2}\rangle$  
is $\mathcal O(L^{1/2})$ with the proportionality constant equal to $\sqrt{\rho c }$. No $p$ dependence is present here. 
Also, we note that  for small enough $p$,  
$c$ is very close to $\rho$ as the second bump peaks at $\rho L$ very sharply. 
Hence in the limit $p \to 0$,  $ \sqrt {\langle \Delta n^2}\rangle $ is given
by  $\rho L^{1/2}$. 

On the other hand, in the fluid phase,  the fluctuation should be finite and hence should either remain 
constant or decrease with $L$. 
 We indeed note from the simulation results that   $\sqrt{\langle \Delta n^2\rangle}$ increases with $L$ for  small $p$ values while for large values of $p$, 
 $\sqrt{\langle \Delta n^2\rangle}$ is independent of  $L$, shown in Fig.~\ref{sr_lr_fluc}(a).
We also find that  $\sqrt{\langle \Delta n^2\rangle}$  shows a  general variation as $\sim L^{\alpha}$ for a fixed value of $p$.
 The value of the exponent  is almost constant, $\alpha\sim 0.50$, up to a 
 value of $p$,  then it sharply decreases to zero with $p$  (see inset of Fig. \ref{sr_lr_fluc}(a)).
  One can estimate the transition point $p_c$  as  the point  (approximately) up to  which 
 the fluctuation varies as $\sim L^{1/2}$ and we get $p_c = 0.33\pm 0.02$ from this data.
The proportionality constant, theoretically predicted as $\rho = 5$ for 
very small $p$, decreases from about $4.8$ at $p=0$  to close to  $4.3$ at 
$p_c$.  
This variation   is not surprising as the value of $c$ 
depends on  $p$.   
Hence we find very good 
agreement 
with the  theoretical prediction
for the   exponent $\alpha$ as well as the associated proportionality constant.  

%
%
%

%

%
%
%
%
\begin{figure}[!h]
\includegraphics[width=8cm,angle=0]{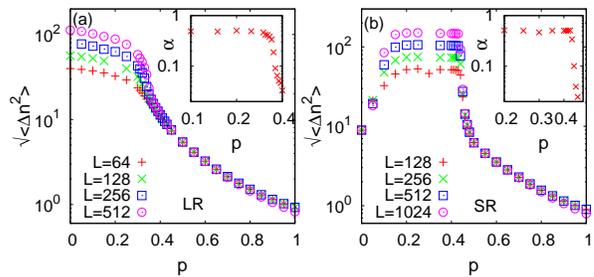}
\caption{Fluctuation of occupation number $\sqrt{\langle \Delta n^2\rangle}$ as a function of $p$ for different  system sizes.
 Inset is for the variation of the exponent $\alpha$  as a function of  $p$.
(a) long range case (LR) (b) short range case (SR). The error bars are less than the size of the data points.
 \label{sr_lr_fluc}
}
\end{figure}

\begin{center}
\begin{table}
\caption{Exponent $\gamma$  for different values of $p$.}
\begin{tabular}{|l|l|l|}
\hline
$p$ & $\gamma$ for $\rho$& $\gamma$ for $\rho$\\
& $=5$ (LR)& $=5$ (SR)\\
 \hline
$0.05$&$0.992\pm0.002$& -\\ 
$0.10$&$0.970\pm0.003$&-\\ 
$0.15$&$0.954\pm0.003$&$ 1.000\pm0.001$ \\ 
$0.20$&$0.940\pm0.005$&$ 1.000 \pm0.001$ \\ 
$0.25$&$0.910\pm0.010$&$0.997\pm0.004$ \\ 
$0.30$&$0.850\pm0.01$&$0.997\pm0.004$ \\ 
$0.35$&-&$0.995 \pm0.005$ \\ 
$0.40$&-&$0.998\pm0.005$ \\ 
\hline

\end{tabular}
\label{tab:expo}
\end{table}
\end{center}

We present snapshots at the steady state in support of the different behaviour of the particle distribution function $P(n)$.
We show snapshots for $p = 0.05$, $0.45$ and $0.70$.
In the first case, there is a condensate forming; even though there are 
a few sites with finite occupancy, the occupancy of the most populated site is order of magnitude higher than the others (Fig.~{\ref{snap_lr}(a)).  
For $p_c < p < 0.5 $, there are no such unique site indicating the absence of the condensate, 
however, the fluctuation in the occupancy is appreciable as seen from  Fig.~{\ref{snap_lr}(b).
As $p \to 1$, all  sites are  occupied with almost equal number 
($n \sim \rho$) 
 (Fig.~{\ref{snap_lr}(c)).

\begin{figure}[!htb]
\begin{center}
 \includegraphics[width=8.5cm]{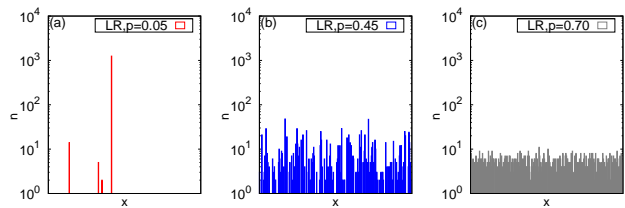}
\end{center}
   \caption{{LR: Snapshots of occupation of sites for different values of $p$. System size is $L=256$ and density $\rho=5$.
   Here  $x$ varies from $1$ to $256$.}}
\label{snap_lr}
\end{figure}

\subsubsection{Short Range (SR) case}
There are some obvious intrinsic differences 
between the long and the short range processes. 
For example, in the extreme case of $p=0$, it is expected that 
many sites will become unoccupied as a rich-gets-richer strategy is imposed. However, the 
condensate will have difficulty to form in the short range model as the sites which are populated may not interact 
at all. Hence a large number of sites, separated by finite  distances, 
become populated though none of them have a macroscopically large occupation number.  As a result,  the distribution
function does not  show a  condensate state. Simulation made  for the short 
range (SR) model shows that this continues up to a value of $p=p_l$. 

\begin{figure}[!h]
\includegraphics[width=7cm,angle=0]{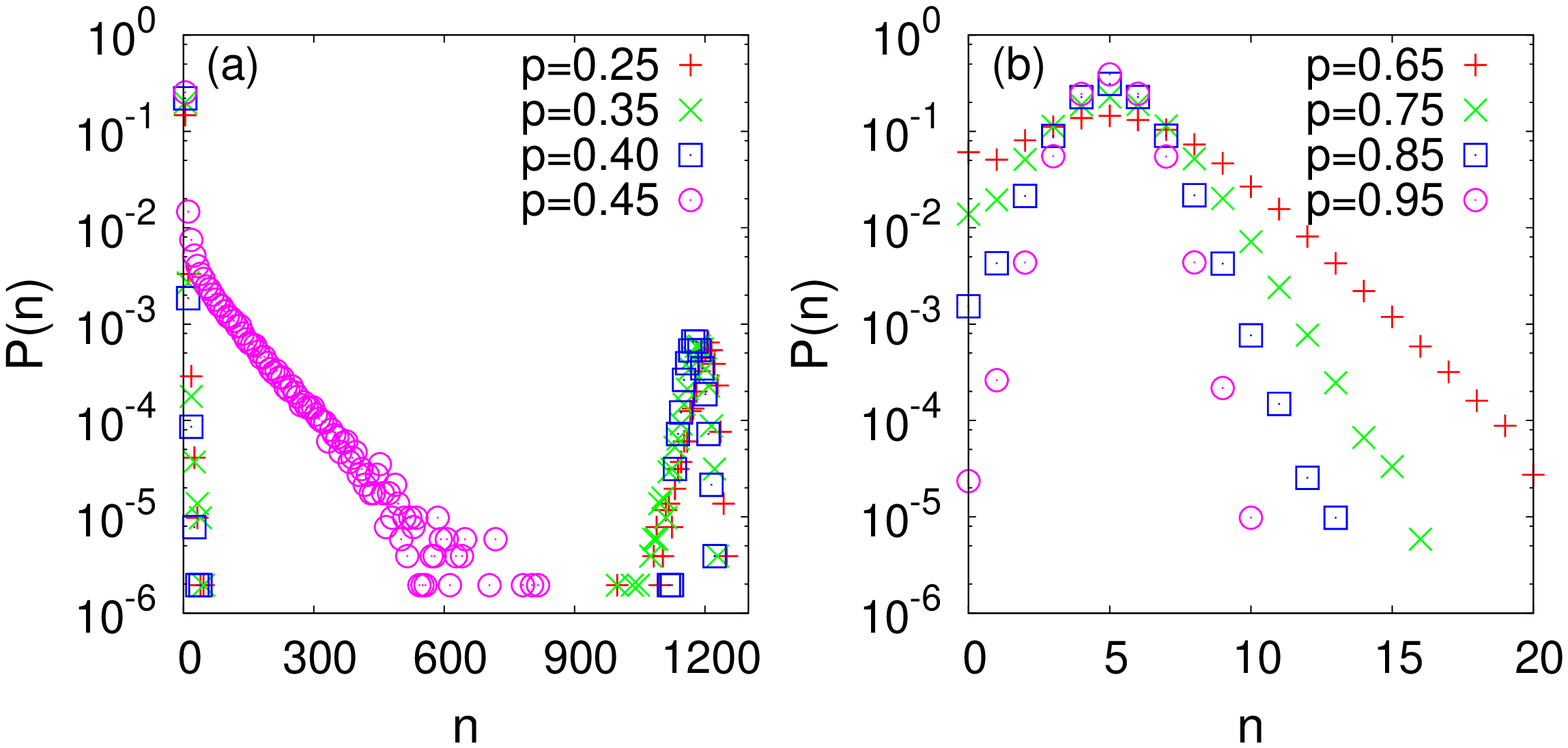}
\caption{SR: Occupation probability distribution $P(n)$ is for (a) lower values of  $p>p_l$ and (b) higher values of  $p$.  System size is $L=256$.}
 \label{sr_dist}
\end{figure}

Hence  in the short range case, only for $p\gtrsim 0.15$ and $p\lesssim 0.44$, 
$P(n)$ shows a bump near $\rho L$  and thus there is  a non-zero lower bound  
of $p = p_l$ 
for the condensate to form (see Fig. \ref{sr_dist}(a) for $P(n)$ at   $p > p_l$;
and section \ref{result}B  for $P(n)$ at $p < p_l$). 
As in the LR case,  
the scaling form (1) is also valid here with
$M \simeq L$ in SR
 (see Fig. \ref{sr_prob_scale}).  However, in contrast to the long range case,  $\gamma$ is very close to unity  and  does not show any 
systematic dependence on $p$ (see Table 1) indicating universal behaviour.  This is explained from the fact that   the secondary peak positions
are almost independent of $p$ except perhaps very close to $p_l \simeq 0.15$. 

\begin{figure}[!h]
\includegraphics[width=8cm,angle=0]{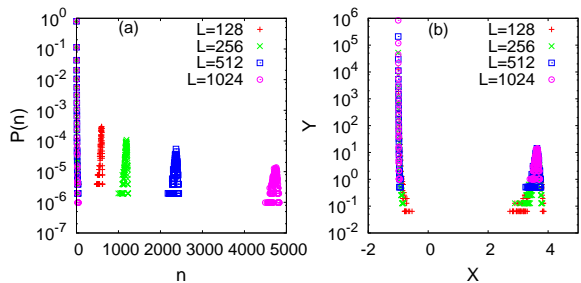}
\caption{SR: (a) Occupation probability distribution $P(n)$ for different  system sizes  for $p=0.35$
and (b) the scaling collapse of the same. Here $Y=P(n) L^{\gamma+1}$ has been plotted against $X=\frac{(n-L)}{L^{\gamma}}$.
Particle density is $\rho=5$.
 \label{sr_prob_scale}
}
\end{figure}

Using the same argument as in the LR case, one  can estimate $p_c$ for the SR case from the data for $\sqrt{\langle \Delta n^2\rangle}$.
For the SR case also, $\sqrt{\langle \Delta n^2\rangle}$  shows system size dependent behaviour for lower values of $p$ only (see Fig.~\ref {sr_lr_fluc}(b)). 
In  Fig. \ref {sr_lr_fluc}(b) (inset), we show that here also $\sqrt{\langle \Delta n^2\rangle}\sim L^{\alpha}$ with $\alpha$ very close to $0.50$ up to a value of $p$
and shows an even
sharper drop beyond this value.
The estimated transition point is $p_c=0.44 \pm 0.01$.
The proportionality constant here is not sensitive to the value of $p$ 
and varies in the range $4.4 \pm 0.2$ (theoretical prediction is $5$).
This is because the secondary peaks in the condensate phase occur more or less at the same position 
independent of $p$ in the SR case. 
Hence once again good agreement is obtained with the theoretical estimates. 

We also present snapshots in support of  the appearance of the variety of phases.
At low values of $p$,  ($p<p_l$), occupied sites cannot interact as they are separated by empty 
sites. The snapshot shows occupancy at many sites, mostly separated by one or a few empty sites (Fig.~{\ref{snap_sr}(a)).
 Hence $P(n)$ does not show condensation for $p< p_l$ but may show a long tail (details in section \ref{result}B). 
 In the condensed phase $p_c>p>p_l$, a finite fraction of
particles lies in a single site (Fig.~{\ref{snap_sr}(b)). 
Immediately above $p_c$, the melting of the condensate state is manifested 
 as we see the majority of particles   populating a cluster of 
 sites slowly disappear in the bulk as $p$ is increased (Figs.~{\ref{snap_sr}(c)-\ref{snap_sr}(e)). 
For even higher value of $p$, all sites are almost equally occupied ($n \sim \rho$) as in LR.

\begin{figure}[!htb]
\begin{center}
 \includegraphics[width=8.5cm]{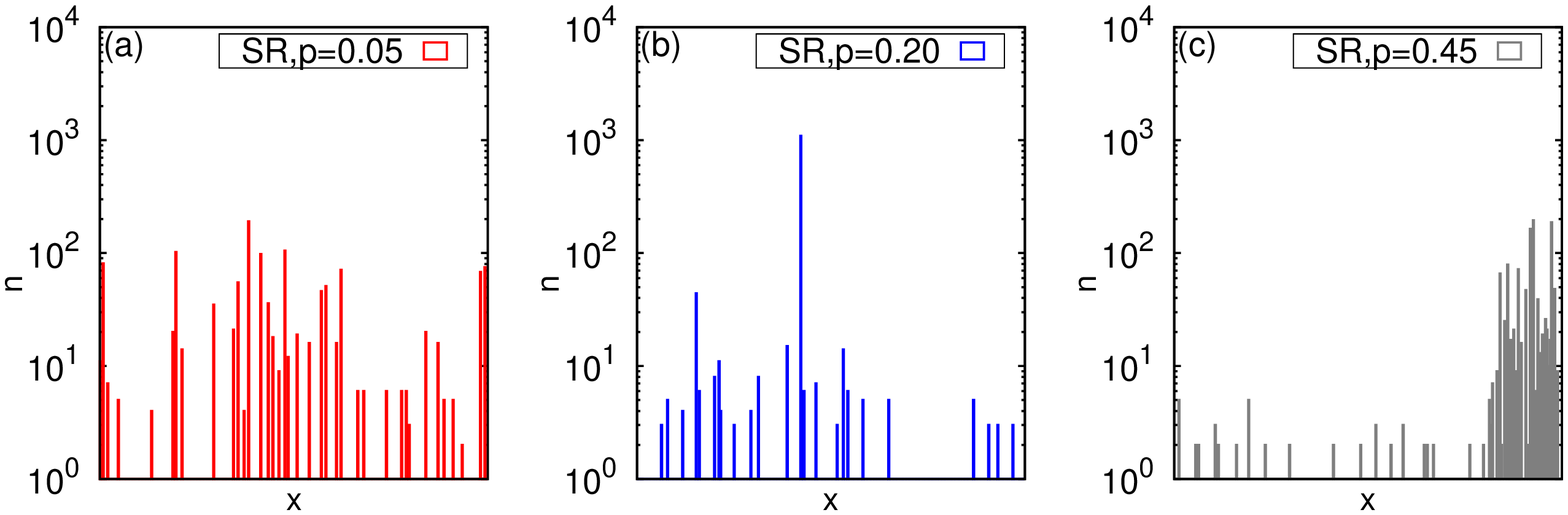} 
 \includegraphics[width=8.5cm]{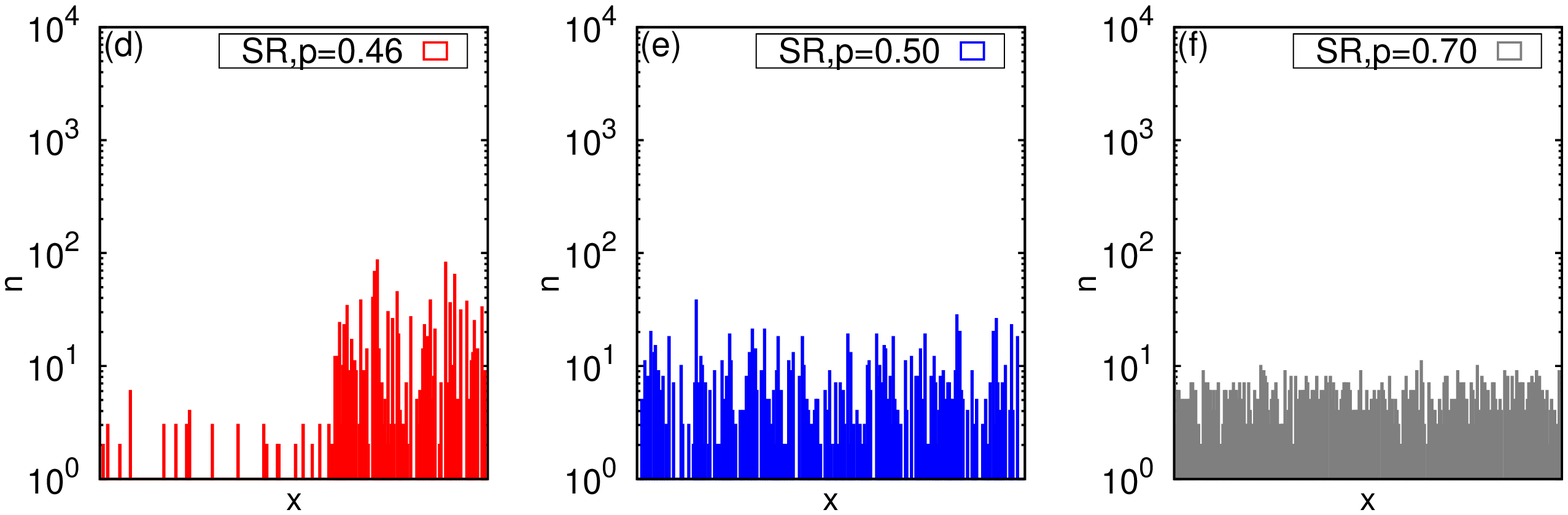}
\end{center}
   \caption{{SR: Snapshots of occupation of sites for different values of $p$. System size is $L=256$ and density $\rho=5$.
   Here $x$ varies from $1$ to $256$.} }
\label{snap_sr}
\end{figure}
\subsubsection{Signature of the phase transition point in other quantities}

\begin{figure}
\begin{center}
 \includegraphics[width=8.5cm]{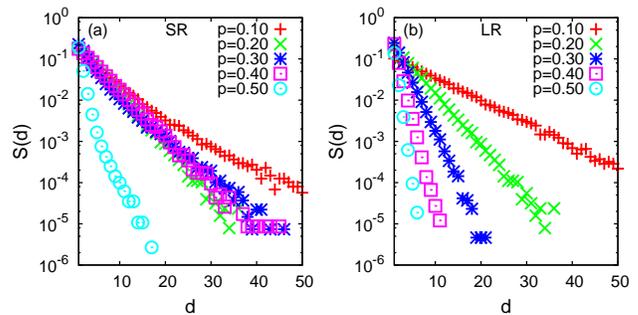}
\end{center}
   \caption{{Distribution of interval size $S(d)$ for different values of $p$.
   Left panel is for the short range model and right one is for the long range model. System size is $L=256$ and density $\rho=5$.}}
\label{size}
\end{figure}

\begin{figure}[!h]
\includegraphics[width=8.5cm,angle=0]{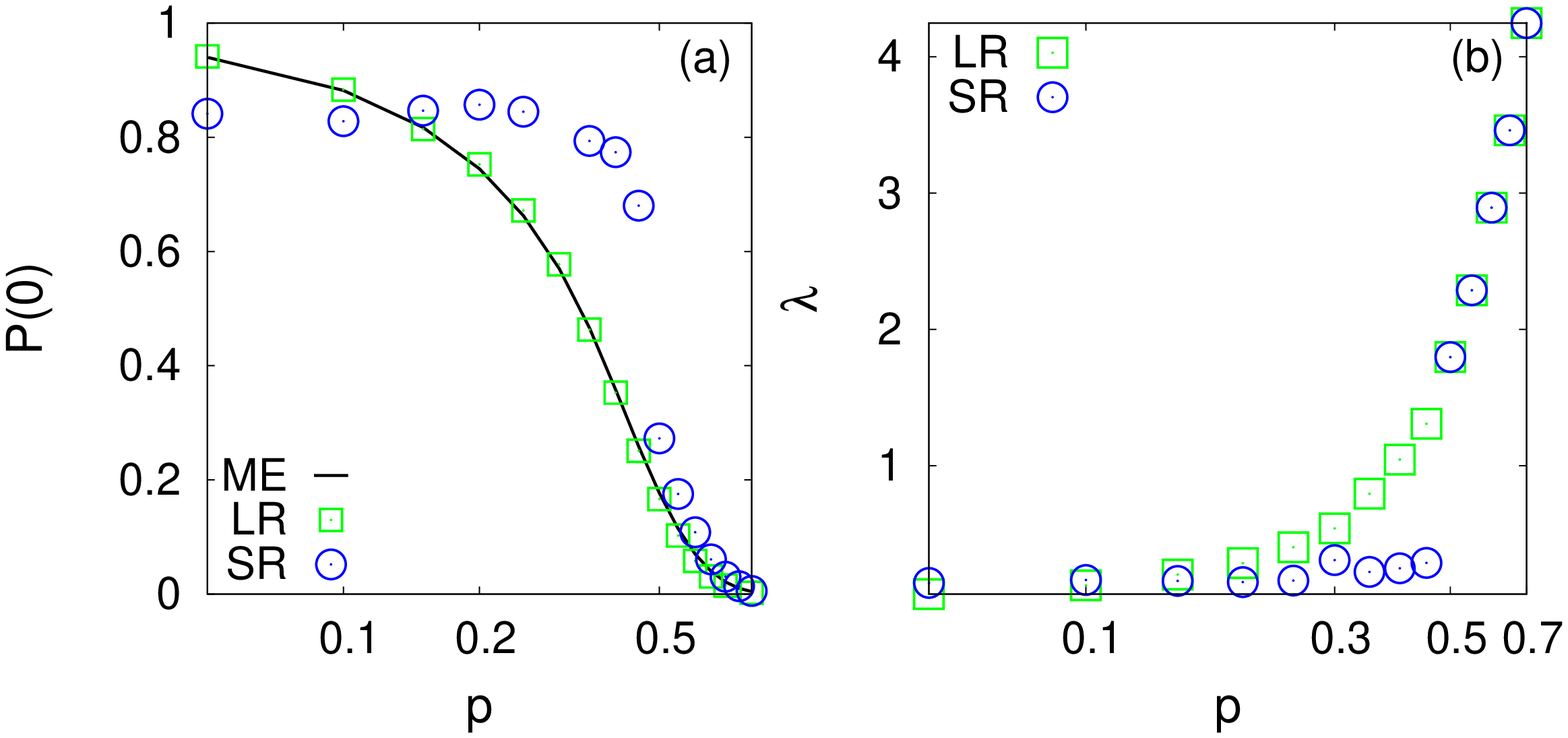}
\caption{(a) Fraction of empty sites $P(0)$ as a function of $p$ for ME, LR and SR methods. (b) Variation of 
decay exponent of separation of two occupied sites $\lambda$ with $p$ for LR and SR methods.}
 \label{empty}
\end{figure}

The signature of the transition point is present in the variation of several quantities in the short range case.
 The distribution $S(d)$ for the interval size $d$  between two occupied sites is  calculated
 which shows an exponential decay: $S(d) \propto  \exp(-\lambda d)$ for all $p$ (Fig.~{{\ref{size}). 
The  decay becomes sharper  as $p$ increases. $S(d)$ has been fitted with the form $\sim \exp (-\lambda x)$ and $\lambda$
has been estimated for all values of $p$. 
The exponent $\lambda$ increases  continuously with $p$ for LR but for SR
$\lambda$ is found to remain rather small up to $p_c$  and it 
increases rapidly beyond this point in the SR model (see Fig. \ref{empty}(b)). This corresponds to having large 
intervals below $p_c$. Few occupied sites are found for  $p_l < p < p_c$ and the 
variation of $\lambda$ with $p$ is a quantitative verification of that.

In the condensate phase, number of empty sites
is appreciably large  while as $p\rightarrow 1$,  all sites become occupied.
Fraction of empty sites $P(0)$  in the steady state has been studied for 
both the long and  short range cases (see Fig. \ref{empty}(a)). 
An estimate is also available from the ME approach which gives steady state values for all $p$ for small $n$. 
 No special behaviour 
is noted in $P(0)$ at $p_c$ in the long range model. 
On the other hand, the short range case clearly shows a jump in  $P(0)$ at
 $p_c$. 
Once again the long range model, expectedly,  does not show any special behaviour. 

\subsection{Nature of  $P(n)$ in different phases: a comparative picture}
In this subsection we present a  detailed study of the functional  behaviour of the distribution  $P(n)$ in different phases.
The behaviour of $P(n)$ in the condensate phase  ($p_l < p <  p_c$) has already been discussed.
Exactly at $p_c$, the behaviour is expected to be a power law. Since numerical estimate of $p_c$ cannot be   exact,
one can only verify this for $p$ close to $p_c$ and a reasonably good agreement is indeed found.    
The entire region $p> p_c$ is a fluid phase. In the short range case, one may 
also  interpret the $p < p_l$ region as a fluid phase as the condensate is absent.  However, at $p=0$ 
the system enters an absorbing (or frozen) state  and even for $< 0 < p < p_l$, the dynamics are quite restricted. We thus identify this region as a 
pseudo frozen phase.  We note the tendency of developing a bump at larger values of $n$ as $p \to p_l \simeq 0.15$ (Fig.~{\ref{fluid}).

\begin{figure}
\begin{center}
 \includegraphics[width=5.5cm]{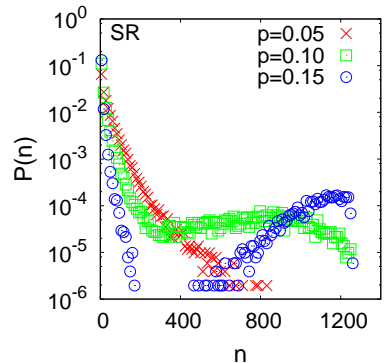}
\end{center}
   \caption{{SR: Distribution of occupation number $P(n)$ for $p\leqslant p_l$ for SR.
   System size is $L=256$ and density $\rho=5$.}}
\label{fluid}
\end{figure}

For $0.5 > p > p_c$, there is a considerable number of sites which are near empty. In the SR case, an interesting
feature is noted: the occupied sites form a cluster which vanishes gradually as $ p \to 0.5$ (see snapshots in Figs.~{\ref{snap_sr}). 
This is a kind of   ``melting" of the highly occupied site 
which exists  up to  $p= p_c$.
The LR case shows no such features for $p > p_c$; naturally 
in the globally connected case, such local correlations  cannot exist.

\begin{figure}[h]
\begin{center}
 \includegraphics[width=8.5cm]{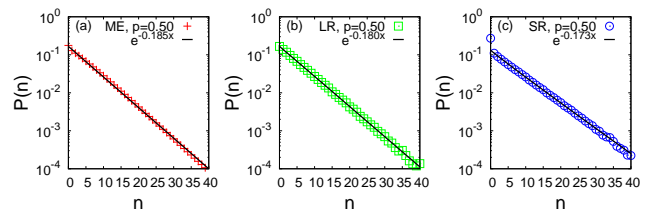}
\end{center}
   \caption{{Distribution of occupation number $P(n)$ for $p=0.5$ for (a) ME, (b) LR and (c) SR. System size is $L=256$ and density $\rho=5$.}}
\label{expo}
\end{figure}

 Both in the long and short range cases, for $p > 0.5$, 
$P(n)$ shows the presence of a  peak (i.e. a most probable value) indicating the emergence of a scale. 
In fact for $ p > 0.5$, the two models give almost identical results  and the results from ME being
more reliable as $p$ increases, also show very good agreement.
Between $p = 0.5$ and 1, $P(n)$ interpolates between a perfectly exponential  and a 
Gaussian form. 

\begin{figure}[!htb]
\begin{center}
 \includegraphics[width=8.5cm]{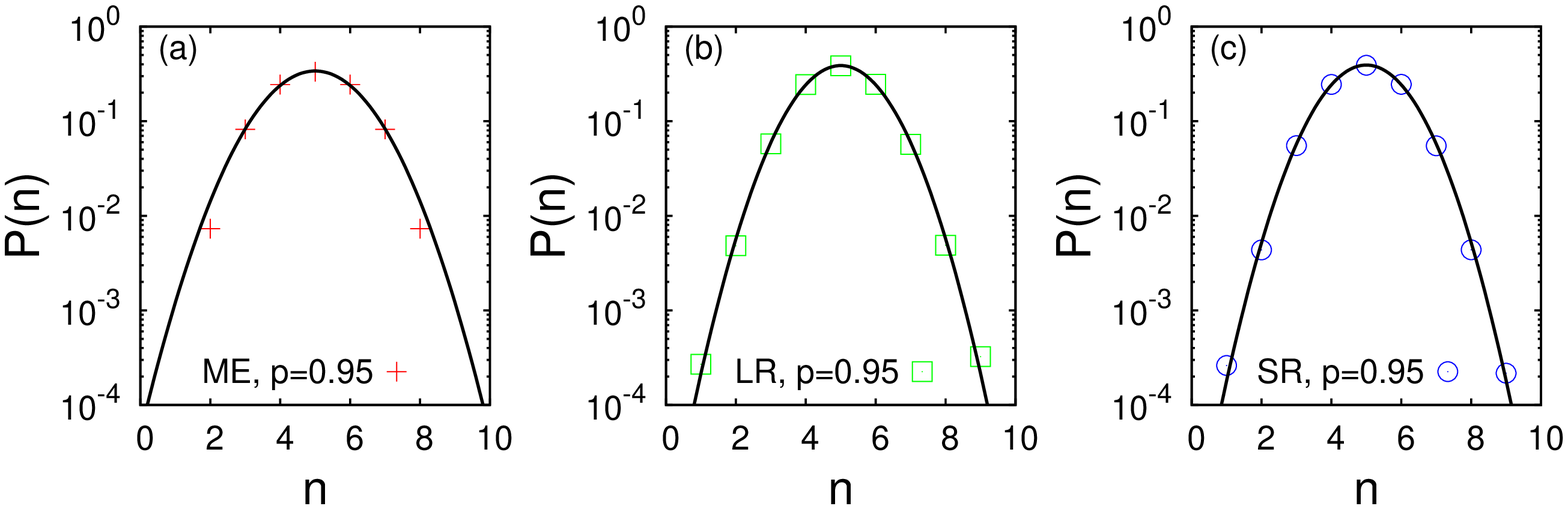}
\end{center}
   \caption{{Distribution of occupation number $P(n)$ and fitted curve using equation \ref{eqn} 
for $p=0.95$ for (a) ME, (b) LR and (c) SR. System size is $L=256$ and density $\rho=5$.} }
\label{gauss}
\end{figure}

$P(n)$ decreases   exponentially with $n$ at $p=0.5$  for all the three cases (Fig.~{\ref{expo}). The fitting form is $A\exp(-bx)$.
The exponents are (1) ME: $A = 0.169$, $ b = 0.185$; (2) LR: $A = 0.166$, $b = 0.181$; (3) SR: $A = 0.119$, $b = 0.154$. Hence although
the qualitative behaviour is independent of the range, the value of the exponent $b$ is case sensitive.
We also note the very good agreement of the results obtained from ME and LR.

For large values  of $p$ ($p>0.5$), $P(n)$ can be fitted to a Gaussian form for ME, LR and SR methods (Fig.~{\ref{gauss}).
 Fittings show better agreement  as $p\rightarrow 1$ and width of the distribution decreases with $p$. 
 The Gaussian form is:
\begin{equation}
 f(x)\sim e^{-a(n-\tilde{n})^2}.
\label{eqn}
\end{equation}
The value of $a$ and $\tilde{n}$ are listed in Table II.

\begin{figure}
\begin{center}
 \includegraphics[width=8.5cm]{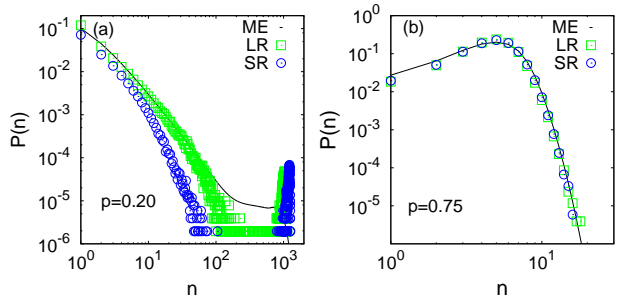}
\end{center}
   \caption{{(a) Distribution of occupation number $P(n)$ for $p=0.20$.
   (b)  Distribution of occupation number $P(n)$ for $p=0.75$. System size is $L=256$ and density $\rho=5$.}}
\label{value}
\end{figure}

\begin{center}
\begin{table}
\caption{ Table for the estimates of the values of the exponents $a$  and $\tilde{n}$ using equation (\ref{eqn}).}
\begin{tabular}{|l|l|l|l|l|l|l|}
\hline
$p$ & $a$ for & $\tilde{n}$ for & $a$ for  & $\tilde{n}$ for  &$a$ for & $\tilde{n}$ for \\
 &  ME & ME & LR &  LR &SR &  SR\\
 \hline
$0.55$&$0.014$&$0.892$&$0.019$&$1.910$&$0.013$&$0.664$ \\ 
$0.60$&$0.034$&$3.984$&$0.043$&$4.215$&$0.023$&$ 3.523$\\ 
$0.65$&$0.060$&$4.682$&$0.077$&$4.761$&$0.058$&$ 4.777$\\ 
$0.70$&$0.092$&$4.895$&$0.121$&$4.926$&$0.105$&$4.972$\\ 
$0.75$&$0.128$&$4.967$&$0.177$&$4.982$&$0.163$&$5.010$ \\ 
$0.80$&$0.169$&$4.990$&$0.244$&$4.991$&$0.231$&$5.007$ \\ 
$0.85$&$0.214$&$4.997$&$0.316$&$5.000$&$0.307$&$5.007$ \\
$0.90$&$0.259$&$4.999$&$0.391$&$5.000$&$0.390$& $5.007$\\ 
$0.95$&$0.306$&$4.999$&$0.468$&$5.000$&$0.480$&$ 5.007$\\ 
$1.00$&$0.352$&$4.999$&$0.541 $&$5.000$&$0.546 $&$ 5.000$\\  \hline
\end{tabular}
\label{tab:expo1}
\end{table}
\end{center}

In the condensed phase, $P(n)$ is quite different for the LR and SR cases.  As mentioned before, the secondary peak of the distribution occurs very close  to $n=\rho L$
for SR model while for LR model, the peak position changes to lesser values of  $n$  
as $p$ approaches $p_c$. 
Fig.~{\ref{value}(a) shows that the actual values of $P(n)$ for LR and SR differ appreciably even for small $n$ for a fixed value of $p$ in the condensate phase. 
The results from the ME approach, as mentioned before, does not show 
perfect   agreement with the  LR results at larger values of $n$ due to computational difficulties at low $p$ values.
But in the fluid phase, there is very good agreement of the values of $P(n)$ for all the three methods (Fig.~{\ref{value}(b)).

\section{Summary and discussions}
\label{summary}

In summary, we have studied  a general form of the the well known zero range process. 
A condensate phase is obtained keeping the density constant and tuning the 
hopping probability $p$. 
A variety of rich behaviour is observed as  $p$ is increased;
the system passes through
different phases for $0 \leq  p < p_l$, $p_l < p \leq  p_c$, $p_c < p < 0.5$ and beyond $p=0.5$. 
Severals snapshots are presented which show the system configuration in the different phases.

Naively one would expect $p_c$, the point below which a condensate phase forms, to be equal to $0.5$ as  hopping from 
lesser populated sites are more probable below $p=0.5$. However we get $p_c< 0.5$ for both the SR and LR models. This is due to the reason
that when an  empty site is involved, the other site will always get less populated, which makes the probability of forming the condensate lesser. Why the 
short range process has a higher $p_c$ compared to the long range case may be explained from the fact that
when an occupied site becomes isolated in the former, no more transfer takes place from it. 

 Our results are  mainly based on finite size scaling analysis. The choice of 
parameters for the data collapse of $P(n)$ is a tricky question and we have taken the values which give the manifestly best collapses. In particular,  
if the value of the 
exponent $\gamma$ is close to unity  the data collapses for the secondary bump in $P(n)$ are not  
sensitive to the  value of $c$ for obvious reasons. However, 
for the long range case, $\gamma$ deviates from unity at larger values of $p$ and the best collapses are obtained for $c=1$ in general. Hence, we have used $c=1$ everywhere
in the plots of the data collapse.
 
The transition points as well as the exponent associated with the scaling function for the particle number distribution function depend on the range of interaction. 
Only the short range case shows universality. 
Excellent agreement is reached for the analytical  and simulation methods (for LR) when  reliable steady state values can be obtained
while  solving the master equation numerically.  
The model shows  range independent behaviour for $p > 0.5$ where a typical scale emerges. 
 $P(n)$ is  exponential for $p=p_c$ while it can be approximated by  a Gaussian form deep inside the fluid phase.

{Acknowledgements}:
AK acknowledges financial support from UGC sanction no. F.7-48/2007(BSR). PS acknowledges financial support from CSIR (Govt. of India) 
project. The authors thank Satya N. Majumdar  for very helpful discussions and suggestions. 
{}

\end{document}